\documentclass[final,5p,times,twocolumn]{elsarticle}

\usepackage[T1]{fontenc}
\usepackage{amssymb}
\usepackage{amsmath}
\usepackage{subcaption}
\usepackage[colorlinks=true,linkcolor=blue,citecolor=blue,urlcolor=blue]{hyperref}

\biboptions{sort&compress}

\journal{Physics Letters B}

\DeclareRobustCommand{\orcid}[1]{\href{https://orcid.org/#1}{\includegraphics[width=8pt]{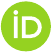}}}

\makeatletter
\let\els@appendix\appendix
\def\appendix{\els@appendix
  \gdef\thesection{\@Alph\c@section}%
  \def\@seccntformat##1{\appendixname~\csname the##1\endcsname. }%
  \@addtoreset{figure}{section}\@addtoreset{table}{section}%
  \setcounter{figure}{0}\setcounter{table}{0}}
\makeatother

\newcommand{\Msun}{\ensuremath{M_\odot}}
\newcommand{\Mtot}{\ensuremath{M_{\rm tot}}}
\newcommand{\FAR}{\ensuremath{\mathrm{FAR}}}
\newcommand{\NCAND}{48}       
\newcommand{\NDET}{47}         
\newcommand{\NDETRUNS}{9, 4, 15, and 19}
\newcommand{\NPOP}{269}
\newcommand{\NPOPHL}{232}
\newcommand{\NCWB}{120}
\newcommand{\NMADCWB}{44}
\newcommand{\MEDMTOTMAD}{69}

\newcommand{\MTOTMIN}{14}
\newcommand{\MTOTMAX}{236}
\newcommand{\SNRMIN}{11.1}
\newcommand{\VTPEAK}{9.9}      
\newcommand{\VTPEAKOTHREE}{3.7}
\newcommand{\RHOMIN}{5}        
\newcommand{\LIVEMIN}{266}
\newcommand{\LIVEMAX}{623}
\begin{document}

\begin{frontmatter}

\title{\textbf{MADGRAV}: a multilevel anomaly-detection pipeline for gravitational-wave searches applied to LIGO data}

\author[1]{Gianluca Inguglia\,\orcid{0000-0003-0331-8279}\corref{cor1}}
\ead{gianluca.inguglia@oeaw.ac.at}
\author[1]{Huw Haigh\,\orcid{0000-0003-1567-0907}}
\ead{huw.haigh@oeaw.ac.at}
\author[1]{Ulyana Dupletsa\,\orcid{0000-0003-2766-247X}}
\ead{ulyana.dupletsa@oeaw.ac.at}
\author[2,3]{Alessandro Longo\,\orcid{0000-0003-4254-8579}}
\ead{alessandro.longo@uniurb.it}
\ead{alessandro.longo@fi.infn.it}

\cortext[cor1]{Corresponding author}

\affiliation[1]{organization={Marietta Blau Institute for Particle Physics, Austrian Academy of Sciences},
            city={Vienna},
            country={Austria}}
\affiliation[2]{organization={Universit\`a degli Studi di Urbino `Carlo Bo'},
            city={I-61029 Urbino},
            country={Italy}}
\affiliation[3]{organization={INFN, Sezione di Firenze},
            city={I-50019 Sesto Fiorentino, Firenze},
            country={Italy}}

\begin{abstract}
We present the results of \textbf{MADGRAV}, a deep-learning-based search for high-mass compact binary coalescences, applied to the data collected by the LIGO interferometers during the third observing run and during the first and second part of the fourth observing run. The \textbf{MADGRAV} pipeline consists of a series of sequential convolutional neural networks that perform anomaly detection, glitch classification, coherence testing, and signal ranking.
Data from the Hanford and Livingston LIGO detectors are studied (both individually and in coherence)  by way of 1 second Q-transform windows. Of the candidates that survive every stage of the pipeline, \NCAND{} reach the significance threshold, and we report \NDET{} gravitational wave detections characterised by a false alarm rate below $1\,{\rm yr}^{-1}$ with a probability of astrophysical origin $p_{\rm astro}>0.9$. Of the \NDET{} detections, \NMADCWB{} are shared with the minimally modelled coherent WaveBurst search. The observed total source-frame masses, extracted from official gravitational wave transient catalogues, are in the $\MTOTMIN$--$\MTOTMAX\,\Msun$ range with a median of $\MEDMTOTMAD\,\Msun$, and a median SNR of 16. We note that the recovered fraction of confident detections rises with mass: for LIGO detectors network SNR $>10$ the pipeline recovers $8.1\%$ of confident catalog events below $30\,\Msun$, $39.8\%$ between $30$ and $100\,\Msun$, and $53.3\%$ above $100\,\Msun$, corresponding to $33.3\%$, $45.5\%$ and $53.3\%$ of the events detected by coherent WaveBurst in the same bins. These results suggest that anomaly detection pipelines can serve as an independent detection channel complementary to matched filtering in the high-mass high-SNR regime.
\end{abstract}


\end{frontmatter}

\section{Introduction}
\label{sec:intro}

Searches for gravitational waves (GWs) from compact binary coalescences (CBCs) in the LIGO-Virgo-KAGRA (LVK) detectors~\cite{LIGOScientificCollaboration2020} can be categorised into two main classes: matched filtering, which is optimal when the signal waveform is known~\cite{10.1119/1.1969250,Allen:2005fk} and with minimally modelled burst searches~\cite{Klimenko:2015ypf,DRAGO2021100678}. The (LVK) catalogs
GWTC-2.1~\cite{LIGOScientific:2021usb}, GWTC-3~\cite{KAGRA:2021vkt}, GWTC-4.0~\cite{LIGOScientific:2025slb,LIGOScientific:2025zdk},
and GWTC-5.0~\cite{LIGOScientific:2026wfs,LIGOScientific:2026sit} combine the template-based
pipelines GstLAL~\cite{2017PhRvD..95d2001M}, PyCBC~\cite{Usman:2015kfa}, and MBTA~\cite{Aubin:2020goo} with
the minimally modelled coherent WaveBurst (cWB) pipeline~\cite{Klimenko:2015ypf,DRAGO2021100678}.
For sufficiently high detector-frame masses~\cite{LIGOScientific:2021tfm} the observable signal may shrink to a few cycles in the sensitive band of current GW observatories, and the presence of possible features that may be incompletely represented in search template banks, such as higher-order modes, precession, eccentricity, provide further motivation for unmodelled or weakly modelled searches. GW190521~\cite{LIGOScientific:2020iuh} and, most recently GW231123~\cite{LIGOScientific:2025rsn} are examples of IMBH production in mergers where cWB had a very important role in their detection.

Several machine learning searches have also been developed, for example~\cite{George:2016hay,Gabbard:2017lja,Schafer:2022dxv,Marx:2024wjt,Chan:2019fuz}; while most utilise waveforms templates directly in their training sample and therefore partially inherit the model dependence of matched filtering, unsupervised and semi-supervised alternatives have also been explored; examples include convolutional and
recurrent autoencoder anomaly searches~\cite{Morawski:2021kxv,Moreno:2021fvp,Raikman:2023ktu,Inguglia:2025cig}, the machine learning burst pipeline MLy~\cite{Skliris:2020qax}, autoencoders trained exclusively on simulated noise~\cite{Guo:2025nxg}, and template-free searches that use inter-detector coincidence as the training objective itself~\cite{Ratner:2026uyt}. Here we present \textbf{MADGRAV}, a Multilevel Anomaly Detection pipeline for GRAVitational wave science. \textbf{MADGRAV} takes the anomaly detection route and further develops the model proposed by the authors in~\cite{Inguglia:2025cig}: its core discriminator is a convolutional autoencoder (CAE) whose reconstruction behavior separates candidate astrophysical transients from detector noise, with a minimal template bank used as a means of weak supervision to increase the discriminating power of the autoencoder to flag potential signal of interest as anomalies with respect to typical detector noise. \textbf{MADGRAV} originates from the anomaly-detection pipeline we introduced for Einstein Telescope mock data in Ref.~\cite{Inguglia:2025cig}. The pipeline presented herein adapts that original work to real data from the LIGO network, and includes additional layers of convolutional neural networks (CNNs) to suppress glitch contamination and implements detector coherence. Section~\ref{sec:pipeline} describes the full architecture and training, Section~\ref{sec:method} introduces the blind search performed on LIGO data, and the results are shown and compared to other pipelines in Sections~\ref{sec:results} and~\ref{sec:comparison}, respectively. Section~\ref{sec:discussion} provides an outlook for future development and deployment of the pipeline.

\section{The \textbf{MADGRAV} pipeline}
\label{sec:pipeline}

\textbf{MADGRAV} operates on time--frequency representations of whitened strain from
the two LIGO detectors with an architecture derived from the
convolutional-autoencoder pipeline of Ref.~\cite{Inguglia:2025cig}, re-engineered
for the realistic two-detector LIGO network and tested on real LIGO data collected in the third and fourth observing runs, specifically in the so-called O3a,b and O4a,b. This work made use of \textsc{gwpy}~\cite{MACLEOD2021100657}, \textsc{PyTorch}~\cite{2019arXiv191201703P},
\textsc{NumPy}~\cite{2020Natur.585..357H}, and \textsc{SciPy}~\cite{2020NaMet..17..261V}, and all codes to reproduce the results are available  in the public \textbf{MADGRAV} repository~\cite{madgrav_repo}.

\subsection{Architecture and training}
\label{sec:architecture}
The encoder component of the CAE consists of three convolutional blocks ($3\times3$ kernels; 32, 64, and 128 filters), each followed by batch normalization, ReLU activation, dropout ($p=0.2$), and $2\times2$ max-pooling whose switch indices are retained. The flattened latent space has
$128\times32\times16=65{,}536$ dimensions. The latent space is deliberately overcomplete: the separation between signal and noise reconstruction errors is produced by the margin objective[Eq.~(\ref{eq:margin})]. Appendix~\ref{app:bottleneck} shows that compressing the latent space by up to a factor of four leaves the detections and the injection efficiency unchanged at fixed false-alarm rate (FAR), and that stronger compression degrades them. The decoder mirrors the encoder and max-unpools with the retained switch indices, so the reconstruction is registered pixel-for-pixel with the input, and the detection feature is a reconstruction error (mean squared error) on that registered map. A single linear layer mapping the latent vector to a signal--noise logit is retained from~\cite{Inguglia:2025cig} but enters neither the training objective of the frozen model nor the detection statistic. Following~\cite{Inguglia:2025cig}, the training objective penalises the model when signal-injected spectrograms are reconstructed with an MSE comparable to
that of noise-only 1--s Q-Transform segments (to which we will refer to as tiles). Denoting by $\mathcal{L}_{\rm noise}$ and $\mathcal{L}_{\rm anom}$ the mean reconstruction MSE of noise-only and signal-injected tiles in a batch, the loss is
\begin{equation}
\mathcal{L}=\mathcal{L}_{\rm noise}
+\lambda\,\mathrm{ReLU}\!\bigl(m\,\bar{\mathcal{L}}_{\rm noise}-\mathcal{L}_{\rm anom}\bigr),
\label{eq:margin}
\end{equation}
where $\bar{\mathcal{L}}_{\rm noise}$ is the noise term held fixed (no gradient)
inside the margin, so that the second term drives the signal reconstruction
error up rather than the noise error down. The ReLU returns zero once
$\mathcal{L}_{\rm anom}\ge m\,\bar{\mathcal{L}}_{\rm noise}$, i.e.\ once signals
reconstruct at least $m$ times worse than noise. We use $m=3$ and $\lambda=2$.

The per-window reconstruction error is calculated and standardised with respect to statistical fluctuations expected in noise-only samples. These per-detector significance values ($\sigma_{\rm H}$, $\sigma_{\rm L}$) constitute the anomaly score of a given window. The CAE is trained in two stages: ten epochs of unsupervised reconstruction of O3a detector noise (MSE on noise-only tiles), followed by ten epochs of fine-tuning with the margin objective of Eq.~(\ref{eq:margin}) on paired noise-only and signal-injected tiles. Injections are drawn from a bank of binary black hole merger waveforms
($102,400$ waveforms using as an approximant IMRPhenomPv2~\cite{Hannam:2013oca}) with primary source masses of $10$--$120\,\Msun$, mass ratios up to 6, luminosity distances $100$--$5000$\,Mpc, and isotropic sky locations and orientations, projected onto both detectors and rescaled to network SNR $\rho_{\rm net}$ uniform in $[8,25]$. The model is trained using uniquely coincident O3a LIGO Livingston and Hanford data that satisfy CAT1 and CAT2 data-quality vetoes~\cite{LIGO:2021ppb,LIGOScientific:2021usb} and carry no hardware injection flag. We randomly select 32 of the 143 available $4096$\,s blocks, giving $36.4$\,h per detector. For each detector, the strain is whitened against an amplitude spectral density (ASD) estimate, we select whitened segments of 4\,s length and their central 2\,s window region is transformed with a multi-resolution Q-transform~\cite{Chatterji:2004qg} ($Q\in[4,64]$, 10--1291\,Hz), of the 2\,s Q-transform region we select the central 1\,s resampled onto a $256\times128$ frequency--time tile and min--max normalized for further analysis (these 1\,s Q-Transform segments are used for training and are the data the full pipeline works with). The final zoom to $256\times128$ reduces each tile by a factor $\sim$39 in pixel count ($0.5\,\mathrm{Hz}\times2.0\,\mathrm{ms}\rightarrow5.0\,\mathrm{Hz}\times7.9\,\mathrm{ms}$), which reduces the computational cost, including that of the time-slide background estimation ($\sim$$10^{9}$ samples), by the same factor. The downsampling was decided based on a measurement: on the recovered
events, whitened noise decorrelates in $\sim$$6.6\,\mathrm{ms}$ while the coherent chirp track remains correlated over $\sim$$12\,\mathrm{ms}$ for $M_{\rm tot}\gtrsim60\,\Msun$, so the $7.9\,\mathrm{ms}$ time bin lies between the two scales and averages down incoherent fluctuations while preserving the signal.
The Q-transform kernel bandwidth also sets a correlation length of $\sim$$14\,\mathrm{Hz}$, leaving the $5.0\,\mathrm{Hz}$ binning oversampled by $\sim$2.8. A future version of \textbf{MADGRAV} can in principle benefit by further optimising frequency and time resolution per mass bin search.

Segmenting the 32 data blocks at 4\,s with a 2\,s stride yields $131{,}008$ tiles, split $60/20/20$ into $78{,}606$ training, $26{,}201$ validation, and $26{,}201$ test tiles. If one of the selected blocks contains a known signal, we remove from that segment a section of $\pm64$\,s around the GPS time of the event (we perform this removal in seven instances). Weak supervision is applied via pairing every noise tile with one injected tile built from the same window; the waveform is drawn uniformly with replacement from the projected bank ($102{,}400$ projections) and placed with a uniform random time offset of up to $\pm0.5$\,s, such that the training and validation splits carry $78{,}606$ and $26{,}201$ injections respectively. We use the Adam optimiser~\cite{adam} (learning rate $10^{-3}$, weight decay
$10^{-5}$, batch size 64, with plateau learning-rate halving) in both stages; the frozen model is the fine-tuning epoch that maximises the number of validation injections whose reconstruction error exceeds the $3\sigma$ point of the validation-noise distribution. The resulting trained model is frozen and utilised throughout the rest of the analysis. The CAE assigns per-detector anomaly scores on the 1\,s tiles over the coincident data; we define the network excess significance of a zero-lag window,
\begin{equation}
\sigma_{\rm net}=\frac{\sigma_{\rm H}+\sigma_{\rm L}}{\sqrt{2}}.
\label{eq:signet}
\end{equation}
A threshold at $\sigma_{\rm net}>4.0$ is set to produce an anomaly detection trigger.

Three auxiliary networks share one compact topology derived from the CAE: four convolutional blocks ($3\times3$ kernels; 16, 32, 64, 128 filters; batch normalization, ReLU, $2\times2$ pooling), global average pooling, and a $128\to64\to1$ fully connected head with dropout (0.3). The first network is the glitch arm which analyses detector tiles individually, and it is trained on bank injections whose 85\,\% is from from the same IMRPhenomPv2 bank as the weak supervision of the CAE and and 15\,\% from a higher mass IMRPhenomXPHM bank (with $\Mtot\in[100,400]\,\Msun$, $q\leq4$), versus $\sigma$-stratified real O3a tiles extracted from the detectors-stream (with $\sigma_{H,L} \in [1,3], [3,6], [6, max]$). This network is deployed as a \emph{five-seed ensemble}: five
networks differing only in random initialization and batch order are averaged in the logit. The tiles of both zero-lag LIGO detectors are then scored, and one seed of the CNN is used as a gradient-weighted class activation mapping (Grad-CAM~\cite{2016arXiv161002391S}) to locate the time where the signal-like evidence peaks in the anomaly score distribution (we find consistency between all seeds and select randomly). After the glitch arm, two ``specialist" CNNs take as input a coincident two-channel H1/L1 tile stack: one restricted in band in the 20--140 Hz region, and the other in 50--500 Hz. The reason for separating the frequency is that different glitches affect different frequencies, and targeting a specific frequency range provides better glitch classification. Both specialists are trained by binary cross entropy on Q-Transform tiles of both LIGO detectors, with injected bank signals as positives (the IMRPhenomXPHM higher-mass bank, detector-frame $M_{\rm tot}$ 100--400 $\Msun$, $q\leq4$, $f_{\min}=20$ Hz, for the low-frequency specialist and the IMRPhenomPv2 lower-mass, $M_{\rm tot}<100 \Msun$, for the high-frequency one), and real time-slid loud coincidences extracted from data with CAE per-detector $\sigma>5$ as negatives, $\mathcal{O}(10^3)$ examples. The low-frequency specialist, therefore, takes as an input the two-channel (Hanford, Livingston) Q-transform tiles in the 20--140\,Hz band, targeting the promotion of short, low-frequency morphologies of massive binaries and the suppression of low frequency glitches; the high-frequency specialist is its 50--500\,Hz counterpart. Both tiles are centered on a common trigger time selected by the Grad-CAM. 

An essential aspect of the search is the cross-detector coherence statistic, which quantifies whether excess power is consistent between sites,
\begin{equation}
C=\frac{2\,\max_{|\tau|\le\tau_{\max}}\bigl|\sum_t x_{\rm H}(t)\,
x_{\rm L}(t+\tau)\bigr|}{\sum_t x_{\rm H}^{2}(t)+\sum_t x_{\rm L}^{2}(t)},
\label{eq:coh}
\end{equation}
where $x_{\rm H,L}$ are the whitened strain series band-limited to a single 20--500\,Hz band in the 1\,s analysis window and $\tau_{\max}=11$\,ms bounds the lag at the inter-site light-travel time. Instrumental transients are incoherent between the detectors, so this statistic anchors the separation of coincident glitches from astrophysical signals. For all signal candidates we evaluate coherence, which will enter in the final ranking statistics

\subsection{Detection flow}
\label{sec:flow}

Gravitational wave detection in \textbf{MADGRAV} proceeds therefore in three subsequent stages. On a 1\,s stride extracted over the coincident data, each detector whitened and Q-Transformed windows are scored once and cached: the CAE anomaly score $\sigma_d$, and the  glitch-arm logit $g_d$ (the five-seed ensemble mean) are computed for every window, before any threshold. Second, the zero-lag Hanford---Livingston windows are formed and the network excess significance [Eq.~\ref{eq:signet}] is required to exceed $\sigma_{\rm net}>4.0$ to trigger the gate. The cross-correlation statistic [Eq.~(\ref{eq:coh})] is evaluated per trigger, completing the feature vector. Finally, a log-likelihood ratio statistic including all information is built
\begin{equation}
\ln\Lambda=\beta_0+\sum_{k=1}^{7}\beta_k\,\frac{x_k-\mu_k}{s_k},
\label{eq:lr}
\end{equation}
from seven per-candidate features $x_k$: the per-detector CAE 
anomaly scores $\sigma_{\rm H}$, $\sigma_{\rm L}$; the two-detector statistic $C$, the lag-maximized cross-correlation of the band-limited whitened series, symmetrically normalized to $[0,1]$ over lags within the light-travel window; the two per-detector spectral centroids, the energy-weighted mean 
frequency over the band-restricted central window of the whitened series; and the two gated glitch-arm scores. The coefficients $\beta$ were fitted by ridge-regularized, class-balanced logistic regression of time-slid off-source pairings against bank injections scored through the full pipeline, with the coherence coefficient constrained non-negative; $\mu_k$ and $s_k$ are the per-fold means and standard deviations of that fit pooled training sample.

A trigger is admitted as a candidate if $\ln\Lambda\geq4.0$. Only then are the specialist arms evaluated: admitted candidates are CNN scored on band-cropped two-detector tiles, at a common time placement taken from the Hanford 
glitch-arm activation map, and must satisfy the the glitch arm gate $g_{\rm net}\geq g_{\min}$ (Sec.~\ref{sec:method}) and glitch specialist gate $\max(s_{\rm HM},s_{\rm LM})>0.5$.

\begin{figure*}[!t]
\centering
\includegraphics[width=\textwidth]{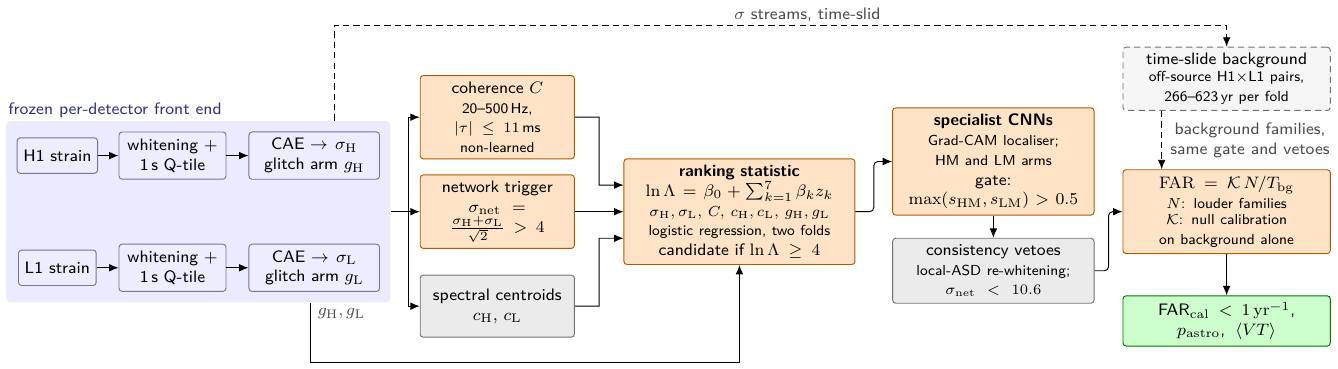}
\caption{Detection flow of \textbf{MADGRAV}. The shaded band is the frozen per-detector front end (whitening, 1\,s Q-tile, autoencoder score $\sigma$, glitch arm $g$); orange boxes are the network-level stages, from the $\sigma_{\rm net}$ trigger through the likelihood-ratio ranking of Eq.~(\ref{eq:lr}) and the gated
specialist networks to the false-alarm rate; grey boxes are non-learned stages. Solid arrows follow a zero-lag candidate; dashed arrows indicate the time-slid background.}
\label{fig:madgrav-flow}
\end{figure*}

Each candidate FAR is then evaluated in the two ranking channels $c\in\{\ln\Lambda,\sigma_{\rm net}\}$; in channel $c$, $N_c$ is the number of time-slide background families of the candidate observing run that pass the same glitch arm and glitch specialist gate, at or above it a its $\sigma_{\rm net}$. The counting applies a local-spectrum consistency veto to candidate (Sec.~\ref{sec:method}). Background triggers sharing the same 4\,s Livingston window across all Hanford partners and slides are counted once (one ``family''). The detection statistic is $\min_c N_c$, converted to a FAR with
a \emph{measured} effective trials factor $N_{\rm eff}$
[Sec.~\ref{sec:method}, Eq.~(\ref{eq:far})]. The statistic of record
(Sec.~\ref{sec:method}) keeps this arm-conditioned counting but drops the $\sigma_{\rm net}$ channel, excludes the zero-lag foreground from the
background, and replaces the asserted factor-4 trials correction under which the candidates were originally selected by an empirically measured calibration factor. Figure~\ref{fig:madgrav-flow} shows a flow chart of the full \textbf{MADGRAV} pipeline.
The complete model (CAE, glitch arms, LR coefficients, and all thresholds) was frozen after the training based on the O3a segments as described above, and applied without retraining to O3b, O4a, and O4b. 

\section{Blind search and significance estimation}
\label{sec:method}

We analyze the public 4\,kHz GWOSC strain~\cite{KAGRA:2023pio,LIGOScientific:2025snk,LIGOScientific:2026jgl} for Hanford--Livingston coincident segments of O3a, O3b, O4a, and O4b. The scanned foreground consists of 725, 556, 818, and 1086 data segments corresponding to  106.8, 96.4, 125.6, and 114.1 days of
coincident livetime, respectively (442.9 days in total). The segment lists are the GWOSC ${\rm H1\_DATA}\cap{\rm L1\_DATA}$ coincident inventories ($\geq16$\,s) of
each run. For all runs the thresholds were fixed in advance and candidate triggers were cross-matched to the catalogs only after the FAR assignment was complete.

The noise background was measured with time slides. Within each observing run, the analysis segments were assigned alternating two indexed subsets (0 and 1). Background coincidences were then formed from time shifted pairings of Hanford and Livingston segments belonging to the same subset. Every candidate was ranked (and its false-alarm rate measured) against the time-slide background of its own subset with the log-likelihood ratio model
fitted only on the other half, so that no candidate or background
trigger is ever scored by a model that has previously seen it. The resulting background livetimes per subset are between $\LIVEMIN$ and $\LIVEMAX$ years: O3a 370 and 405\,yr; O3b 288 and 266\,yr; O4a 495 and 577\,yr; O4b 623 and 520\,yr, for subset 0 and 1, respectively. We restricted the background trigger to time-slide windows satisfying the same $\sigma_{\rm net}>4.0$ condition that defines a foreground trigger. 
The detection statistic minimizes the family rank over the two ranking channels (Sec.~\ref{sec:flow}); the FAR is
\begin{equation}
\FAR=N_{\rm eff}\,\frac{\min_c N_c}{T_{\rm bg}},
\label{eq:far}
\end{equation}
where $N_{\rm eff}$ is the trials factor of the minimum over the two channels. Equation~(\ref{eq:farcal}) below supersedes this form for every quoted number:
the empirical factor $\mathcal{K}$ is measured at trials factor unity and absorbs $N_{\rm eff}$, so the two are never applied together. Here $N_c$ is the integer count of loud background families in the candidate observing run, $T_{\rm bg}$ that fold background livetime, and $N_{\rm eff}$ the effective trials factor of the minimum
\begin{equation}
N_{\rm eff}(x)=\frac{R_{\min}(x)}{\max_c R_c(x)},
\label{eq:neff}
\end{equation}
where $R_{\min}(x)$ is the rate per unit background time at which families reach $\min_c N_c/T_{\rm bg}<x$ under their own leave-one-out ranks, and $R_c(x)$ is the same rate in channel $c$ alone
($N_{\rm eff}=1.86$--$1.97$ with 90\% intervals of $\pm0.3$; O4b fold~1 uses a mildly $x$-dependent value, 1.57 at $1\,{\rm yr}^{-1}$, because the consistency test rejected a single value there (Appendix~\ref{app:robust})). The candidates own zero-lag pair is the only exclusion from its background. For every detection we recover $\min_c N_c$
and attach an exact 90\% Poisson interval~\cite{bdfc2ca9-0182-32da-93ef-5aa1e72bfc40}, the upper limit
replacing the count by $N_{\rm up}=\tfrac12\chi^{2}_{0.90}\bigl[2(N+1)\bigr]$;
candidates with $N=0$ are therefore quoted as 90\% upper limits, $\FAR<\mathcal{K}\times2.30/T_{\rm
bg}$.

\emph{Injection campaigns.} The efficiency, sensitive-volume, and
$p_{\rm astro}$ estimates below use a single per-run injection campaign scored through the identical frozen chain and the adopted background counting. A signal population is drawn in equal parts from three banks: a low-mass bank
(IMRPhenomPv2, non-spinning projections; $\Mtot\in[10,22]\,\Msun$, $q\leq6$),
the stellar bank of Sec.~\ref{sec:architecture} (IMRPhenomPv2, non-spinning projections; $m_1\in[10,120]\,\Msun$, $q\leq6$), and an ultramassive bank (IMRPhenomXPHM, precessing; detector-frame $\Mtot\in[150,400]\,\Msun$, $q\leq4$, lower starting frequency), which together give detector-frame support from $10$ to $400\,\Msun$. Sky position, orientation, and polarization are isotropic; each injection is placed on a network--SNR grid of
$13$ levels spanning $\rho_{\rm net}=5$--$80$, and later importance-weighted to a uniform-in-volume distance prior within the sampled sphere ($\propto\rho^{-4}$; Sec.~\ref{sec:vt}). The grid must span the population at uniform weights: the $\rho^{-4}$ weights are renormalized over the sampled levels, so a ceiling low enough to exclude the loud tail redistributes that weight onto
levels at which the search is inefficient and biases $\langle VT\rangle$ low. The campaigns comprise $122\,850$, $70\,200$, $58\,500$, and $70\,200$ scored injections in O3a, O3b, O4a, and O4b, respectively---$5\,850$ per host segment on $21$, $12$, $10$, and $12$ segments; $307\,226$ of the $321\,750$ survive the horizon-validity cut. The specialist arms were trained on Q-transform crops of the stellar and ultramassive banks against O3a background triggers, and the glitch arm on bank injections versus $\sigma$-stratified real O3a tiles (Sec.~\ref{sec:architecture}).

Surviving candidates are re-whitened against a local $\pm64$\,s Welch ASD estimate; a candidate is vetoed if both localized arm scores fall below 0.5 or, for a network-channel candidate, if $\sigma_{\rm net}$ falls below 4.0 under local whitening. 

Two selection rules were studied using the injection population alone and are applied identically to candidates, to every time-slide background pair and to the injections that measure the efficiency: an upper veto $\sigma_{\rm net}<10.6$ (99\% of gate-passing injections) and a glitch-arm gate $g_{\rm net}\equiv(g_{\rm H}+g_{\rm L})/\sqrt{2}\geq g_{\min}=-4.02$, the first percentile of $g_{\rm net}$ over the $57\,641$ gate-passing, triggering injections of the campaign above ($\le 1\%$ injection loss). The loudest time-slide families are chance coincidences of two short broadband instrumental transients, which the specialist arms score as signal-like as the detections but the glitch arm separates (median $g_{\rm H}/g_{\rm L}=+2.5/+3.3$ for the detections against $-1.6/-3.0$ for $\sim$50 loudest families); the gate passes all \NCAND{} candidates and removes $37\%$ of the background families that set the FARs ($40\%$ of the summed louder-family counts).

We quote FAR as the primary significance and additionally report a per-event probability of astrophysical origin, $p_{\rm astro}$, constructed in direct analogy to the cWB method~\cite{2015PhRvD..91b3005F,LIGOScientific:2025zdk}. Each run signal density is measured from the injection campaign above, pushed through the identical adopted counting that defines the quoted FARs and reweighted to the published-catalog total-mass distribution; the noise density is analytic and pinned to its time-slide null expectation rather than fitted. The construction, fitted mixture parameters, and three systematic checks are given in Appendix~\ref{app:pastro}. 

Every gate-passing, veto-surviving time-slide family is treated as a pseudo-foreground candidate and assigned a $\FAR$ by the identical procedure used for a real candidate, ranked against the remaining background with its own pair excluded. If the quoted $\FAR$ were exact, the number $n(x)$ of such families assigned $\FAR<x$ would equal $E(x)=x\,T_{\rm bg}$; the ratio
\begin{equation}
\mathcal{K}(x)=\frac{n(x)}{E(x)},\qquad E(x)=x\,T_{\rm bg},
\label{eq:kfactor}
\end{equation}
is therefore a direct, foreground-free measurement of how far the quoted rate is from the true one. The test involves no
zero-lag data.

Two properties fix its interpretation. For a plain unconditioned rank the ratio is unity by construction---the $r$-th loudest family has $\FAR=r/T_{\rm bg}$, so exactly $xT_{\rm bg}$ families fall below $x$---and we reproduce $\mathcal{K}=1.00$ for a single-channel rank in all four runs.
And a veto applied consistently to both the candidate and the background leaves a rank unchanged, which we also verify ($1.00$ before and after).

Evaluated on the full pseudo-foreground population of each run, the adopted statistic gives $\mathcal{K}=5.60$, $4.59$, $1.43$ and $3.62$ for O3a, O3b, O4a and O4b.
The dominant contribution is the arm-conditioned counting: because a family is counted only when it beats the candidate on loudness \emph{and} on the candidate own arm score, the count is a conditional tally rather than a rank, and it runs $3$--$13\%$ of the corresponding rank. Taking the minimum over the two channels contributes a further factor measured at $1.9$, close to the value $2$ expected for independent channels and reduced below it only by their correlation. The statistic of record is therefore
\begin{equation}
\FAR_{\rm cal}=\mathcal{K}\,\frac{\min_c N_c}{T_{\rm bg}},
\label{eq:farcal}
\end{equation}
evaluated with the per-run $\mathcal{K}$ above. Two properties of this statistic must be stated explicitly, because they differ from the general form of Eq.~(\ref{eq:far}) and because every number in this paper is computed from them. First, the minimum is taken over the two \emph{arms} and not over the two channels: the $\sigma_{\rm net}$ channel is removed entirely and every candidate is ranked on $\ln\Lambda$ alone, so that
$\min_c N_c \to \min\bigl(N^{\rm HM},N^{\rm LM}\bigr)$. Second, those counts remain conditioned on the candidate own arm scores, a family entering $N^{a}$ only if its arm score exceeds the candidate. $\mathcal{K}$ is measured on exactly this rank function and at trials factor unity, so it already absorbs the $N_{\rm eff}$ of Sec.~\ref{sec:method} and is not applied on top of it.

Every FAR quoted in this paper is $\FAR_{\rm cal}$, and we require it to lie below $1\,{\rm yr}^{-1}$; of the \NCAND{} candidates passing the raw threshold, \NDET{} satisfy this.

\section{Results}
\label{sec:results}
\begin{figure*}[!t]
\centering
\includegraphics[width=\textwidth]{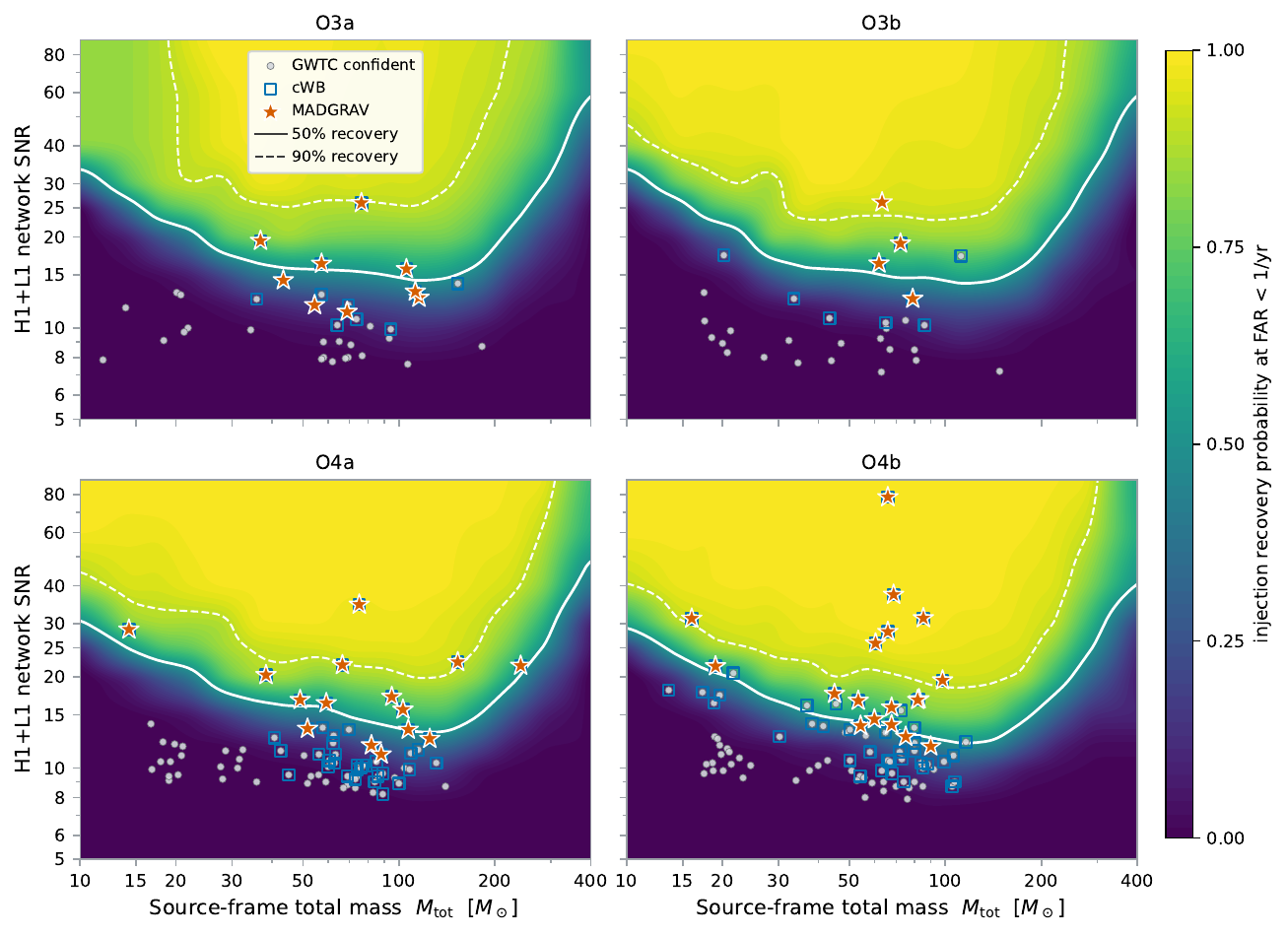}
\caption{Injection recovery over the source-frame total mass---network SNR plane, one panel per run. Colour is the recovery probability at $\FAR<1\,{\rm yr}^{-1}$; white curves are the $50\%$ (solid) and $90\%$ (dashed) contours. Grey circles are GWTC confident H1--L1 coincident events,
blue squares cWB and orange stars \textbf{MADGRAV} detections. The grid used contains $16$ logarithmic mass bins and $13$ SNR levels ($\rho_{\rm net}=5$--$80$) with $\geq40$ injections per cell.}
\label{fig:recovery}
\end{figure*}

\begin{figure*}[!t]
\centering
\begin{subfigure}[b]{0.49\textwidth}
\includegraphics[width=\textwidth]{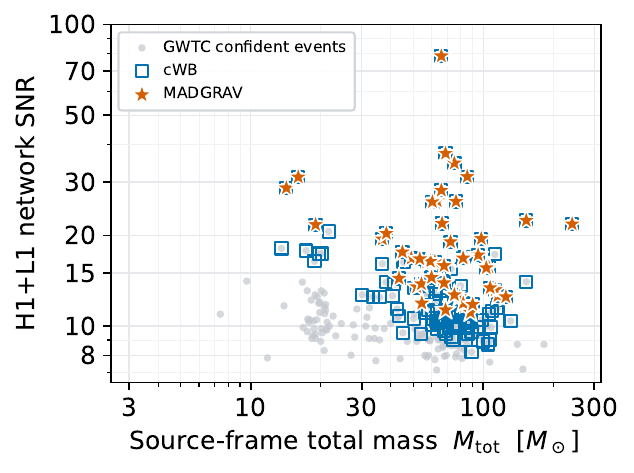}
\caption{}
\label{fig:population}
\end{subfigure}
\hfill
\begin{subfigure}[b]{0.49\textwidth}
\includegraphics[width=\textwidth]{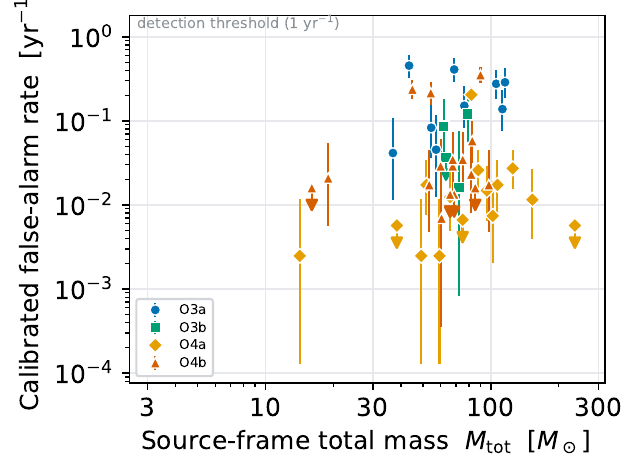}
\caption{}
\label{fig:far}
\end{subfigure}
\caption{\textbf{Left (a):} Source-frame total mass versus H1+L1 network matched-filter SNR for the confident events of GWTC-2.1, GWTC-3, GWTC-4.0, and GWTC-5.0 that were observed by Hanford and Livingston simultaneously (grey), the \NCWB{} identified by cWB at $\FAR<1\,{\rm yr}^{-1}$ (blue squares), and the \textbf{MADGRAV} detections at calibrated $\FAR<1\,{\rm yr}^{-1}$ (orange stars). GW240406\_062847, an event we detected, has no published source-frame total mass~\cite{LIGOScientific:2026wfs} and is not shown in the plot. \textbf{Right (b):} Calibrated FAR with 90\% Poisson intervals versus source-frame total mass for the 46 detections (\NDET{} less GW240406\_062847). Arrows denote 90\% upper limits for the 9 events with no louder background family in their full time-slide set (266--623 yr per observing run).}
\label{fig:pop_and_far}
\end{figure*}

The injection campaign of Sec.~\ref{sec:method} underlies what show in Fig.~\ref{fig:recovery}: each injection is recovered through the same sliding grid, glitch gate and local-ASD veto as a real candidate, and carries the sub-tile placement loss. Figure~\ref{fig:recovery} also shows the detected events per observing run, showing an excellent agreement between expectations and observations. In Fig.~\ref{fig:population} we place the detected events in the catalog mass--SNR plane. Every detection is consistent with a confident event of the published catalogs; and no candidate without a
catalog counterpart survives the analysis.

Table~\ref{tab:eff} gives the resulting efficiency. Recovery reaches $50\%$ near
$\rho_{\rm net}\simeq15$--$20$ and plateaus at $0.86$--$0.97$ rather than at
unity: a residual few per cent of even the loudest injections are lost, so an
extrapolation to perfect recovery at high SNR would be wrong. Comparing
simulation with data at each detection's own cell (Table~\ref{tab:cmp}). 
The search returns \NDET{} detections at calibrated
$\FAR<1\,{\rm yr}^{-1}$: \NDETRUNS{} in O3a, O3b, O4a, and O4b, respectively, with $46.5\pm4.1$ predicted detctions. Table~\ref{tab:detections} lists all \NDET{} with their recovered background counts translated into 90\% intervals and the per-event
$p_{\rm astro}$ (all \NDET{} have $p_{\rm astro}>0.90$ and 42 of them $\geq0.99$). 
All $46$ adopted detections with a published source-frame mass lie inside the measured region; GW240406\_062847 has no published mass.
\begin{table*}[!t]
\centering
\caption{Recovery fraction at ${\rm FAR}<1\,{\rm yr^{-1}}$ versus injected H1+L1
network SNR. A dash marks a level absent for that run. Values are run averages
over the injected mass mix, not per-cell efficiencies.}\label{tab:eff}
\begin{tabular}{lccccccccccccc}
\hline
$\rho_{\rm net}$ & 5 & 6 & 7 & 8 & 10 & 12 & 15 & 20 & 25 & 30 & 40 & 60 & 80\\
\hline
O3a & 0.00 & 0.00 & 0.00 & 0.01 & 0.07 & 0.18 & 0.39 & 0.55 & 0.72 & 0.84 & 0.92 & 0.92 & 0.89\\
O3b & -- & 0.00 & 0.00 & 0.01 & 0.05 & 0.15 & 0.36 & 0.53 & 0.69 & 0.82 & 0.93 & 0.96 & 0.96\\
O4a & 0.00 & 0.00 & 0.00 & 0.00 & 0.04 & 0.12 & 0.32 & 0.52 & 0.73 & 0.83 & 0.92 & 0.96 & 0.97\\
O4b & 0.00 & 0.00 & 0.00 & 0.01 & 0.06 & 0.18 & 0.41 & 0.59 & 0.79 & 0.88 & 0.95 & 0.97 & 0.97\\
\hline
\end{tabular}
\end{table*}

\begin{table}[!t]
\centering
\caption{Detections found versus predicted, evaluated at each event own
$(\Mtot,\rho_{\rm net})$ cell. Uncertainties are Poisson-binomial.}\label{tab:cmp}
\begin{tabular}{lcccc}
\hline
Run & Comparable & Found & Predicted & Difference\\
\hline
O3a & 37 & 9 & $7.8\pm1.8$ & $+0.69\sigma$\\
O3b & 31 & 4 & $4.0\pm1.1$ & $-0.03\sigma$\\
O4a & 76 & 15 & $11.6\pm2.2$ & $+1.54\sigma$\\
O4b & 86 & 18 & $23.2\pm2.7$ & $-1.88\sigma$\\
\hline
Total & 230 & 46 & $46.5\pm4.1$ & $-0.13\sigma$\\
\hline
\end{tabular}
\end{table}

Nine detections have no louder background family in their full
time-slide set and are bounded below $\FAR<0.006$--$0.037\,{\rm yr}^{-1}$ (90\%, Fig.~\ref{fig:far}), the tightest being GW231226\_101520 and GW230927\_153832. These detections include GW250114\_082203 and GW240920\_124024 in O4b, both at $<0.013\,{\rm yr}^{-1}$.
A further 23 detections have $\FAR<0.05\,{\rm yr}^{-1}$. GW231123, the most massive binary black hole reported to date~\cite{LIGOScientific:2025rsn}, is recovered at $\FAR=0.006\,{\rm yr}^{-1}$.
Two different network SNRs are quoted in the catalog literature: the network matched-filter SNR reported by the parameter-estimation analyses, and the network SNR of the triggering search pipeline. For the \NPOP{}
confident events the two differ by more than $2\%$ for 166 of them. Table~\ref{tab:detections} and this section take the parameter-estimation value, from the GWOSC
event API releases of GWTC-2.1, GWTC-3, GWTC-4.0/4.1 and
GWTC-5.0~\cite{LIGOScientific:2021usb,KAGRA:2021vkt,LIGOScientific:2025slb,LIGOScientific:2026wfs}, as the common starting point; the table reports it as published and Fig.~\ref{fig:population} the H1+L1-only value derived from it below, so the two are related by a stated conversion
rather than being two independently sourced numbers.
\begin{table*}[!t]
\centering
\setlength{\tabcolsep}{4.5pt}
\caption{\label{tab:detections}The \NDET{} \textbf{MADGRAV} detections at calibrated $\FAR<1\,{\rm yr}^{-1}$. $M_{\rm tot}$ is the published source-frame total mass and $\rho_{\rm net}$ the published network matched-filter SNR \cite{LIGOScientific:2021usb,KAGRA:2021vkt,LIGOScientific:2025slb,LIGOScientific:2026wfs}, over the full observing network; $\rho_{\rm H1+L1}$ is the H1+L1-only value of Eq.~(\ref{eq:snrhl}), the quantity this coincident search could
reach and the axis of Fig.~\ref{fig:population}. The two differ only where Virgo also observed. $\sigma_{\rm net}$ is the \textbf{MADGRAV} network excess significance. Events with no louder background family ($N=0$) carry a $90\%$ upper limit.
$p_{\rm astro}$ is a per-run Poisson-mixture (FGMC) fit on the foreground-excluded FAR axis.
The last column marks candidates also identified by cWB at ${\rm FAR}<1\,{\rm yr}^{-1}$.}
\begin{tabular}{llcrrrccccc}
\hline 
Event & Run & $M_{\rm tot}$ [$\Msun$] & $\rho_{\rm net}$ & $\rho_{\rm H1+L1}$ & $\sigma_{\rm net}$ & $N$ & $\FAR$ [yr$^{-1}$] & $\FAR_{\rm incl}$ & $p_{\rm astro}$ & cWB \\
\hline
GW190408\_181802 & O3a & 43 & 14.6 & 14.4 & 5.8 & 30 & $0.454$ & $0.454$ & $1.00$ & \checkmark \\
GW190412 & O3a & 37 & 19.8 & 19.4 & 6.1 & 3 & $0.041$ & $0.041$ & $1.00$ & \checkmark \\
GW190513\_205428 & O3a & 54 & 12.5 & 11.9 & 5.9 & 6 & $0.083$ & $0.083$ & $1.00$ &  \\
GW190519\_153544 & O3a & 106 & 15.9 & 15.7 & 7.5 & 20 & $0.276$ & $0.276$ & $1.00$ & \checkmark \\
GW190521\_074359 & O3a & 76 & 25.9 & 25.9 & 8.1 & 10 & $0.151$ & $0.151$ & $1.00$ & \checkmark \\
GW190602\_175927 & O3a & 116 & 13.2 & 12.6 & 8.8 & 19 & $0.288$ & $0.288$ & $1.00$ & \checkmark \\
GW190706\_222641 & O3a & 113 & 13.4 & 13.2 & 8.2 & 10 & $0.138$ & $0.138$ & $1.00$ & \checkmark \\
GW190727\_060333 & O3a & 69 & 11.7 & 11.3 & 6.5 & 27 & $0.409$ & $0.409$ & $1.00$ & \checkmark \\
GW190828\_063405 & O3a & 57 & 16.5 & 16.3 & 6.7 & 3 & $0.045$ & $0.045$ & $1.00$ & \checkmark \\ \hline
GW191222\_033537 & O3b & 79 & 12.5 & 12.5 & 7.3 & 7 & $0.121$ & $0.121$ & $1.00$ & \checkmark \\
GW200129\_065458 & O3b & 63 & 26.8 & 26.0 & 8.5 & 0 & $<0.037$ & $<0.037$ & $1.00$ &  \\
GW200224\_222234 & O3b & 72 & 20.0 & 19.0 & 10.1 & 1 & $0.016$ & $0.016$ & $1.00$ & \checkmark \\
GW200311\_115853 & O3b & 62 & 17.8 & 16.3 & 8.1 & 5 & $0.086$ & $0.086$ & $1.00$ & \checkmark \\
\hline
GW230601\_224134 & O4a & 107 & 13.4 & 13.4 & 6.9 & 6 & $0.017$ & $0.017$ & $1.00$ & \checkmark \\
GW230627\_015337 & O4a & 14 & 28.7 & 28.7 & 6.2 & 1 & $0.002$ & $0.002$ & $1.00$ & \checkmark \\
GW230628\_231200 & O4a & 59 & 16.4 & 16.4 & 6.8 & 1 & $0.002$ & $0.002$ & $1.00$ & \checkmark \\
GW230707\_124047 & O4a & 82 & 11.9 & 11.9 & 5.1 & 71 & $0.205$ & $0.323$ & $0.91$ & \checkmark \\
GW230824\_033047 & O4a & 88 & 11.1 & 11.1 & 4.6 & 9 & $0.026$ & $0.029$ & $1.00$ & \checkmark \\
GW230914\_111401 & O4a & 96 & 17.2 & 17.2 & 7.5 & 6 & $0.015$ & $0.015$ & $1.00$ & \checkmark \\
GW230919\_215712 & O4a & 49 & 16.8 & 16.8 & 4.7 & 1 & $0.002$ & $0.002$ & $1.00$ & \checkmark \\
GW230922\_040658 & O4a & 125 & 12.5 & 12.5 & 5.6 & 11 & $0.027$ & $0.027$ & $1.00$ & \checkmark \\
GW230924\_124453 & O4a & 52 & 13.5 & 13.5 & 5.4 & 6 & $0.017$ & $0.017$ & $1.00$ & \checkmark \\
GW230927\_153832 & O4a & 38 & 20.3 & 20.3 & 5.8 & 0 & $<0.006$ & $<0.006$ & $1.00$ & \checkmark \\
GW231028\_153006 & O4a & 153 & 22.4 & 22.4 & 8.9 & 4 & $0.012$ & $0.012$ & $1.00$ & \checkmark \\
GW231102\_071736 & O4a & 102 & 15.6 & 15.6 & 7.2 & 3 & $0.007$ & $0.007$ & $1.00$ & \checkmark \\
GW231123\_135430 & O4a & 236 & 21.8 & 21.8 & 9.1 & 0 & $<0.006$ & $<0.006$ & $1.00$ & \checkmark \\
GW231206\_233901 & O4a & 66 & 21.9 & 21.9 & 8.2 & 5 & $0.012$ & $0.012$ & $1.00$ & \checkmark \\
GW231226\_101520 & O4a & 75 & 34.7 & 34.7 & 9.2 & 0 & $<0.007$ & $<0.007$ & $1.00$ & \checkmark \\ \hline
GW240514\_121713 & O4b & 83 & 16.9 & 16.9 & 6.8 & 10 & $0.058$ & $0.070$ & $1.00$ & \checkmark \\
GW240615\_113620 & O4b & 60 & 26.4 & 25.8 & 8.9 & 1 & $0.007$ & $0.007$ & $1.00$ & \checkmark \\
GW240621\_195059 & O4b & 66 & 28.2 & 28.2 & 8.5 & 0 & $<0.013$ & $<0.013$ & $1.00$ & \checkmark \\
GW240705\_053215 & O4b & 82 & 16.8 & 16.8 & 6.7 & 4 & $0.023$ & $0.023$ & $1.00$ & \checkmark \\
GW240919\_061559 & O4b & 68 & 16.9 & 15.9 & 7.5 & 5 & $0.029$ & $0.029$ & $1.00$ & \checkmark \\
GW240920\_124024 & O4b & 69 & 37.4 & 37.4 & 8.7 & 0 & $<0.013$ & $<0.013$ & $1.00$ & \checkmark \\
GW240925\_005809 & O4b & 16 & 31.2 & 31.2 & 6.4 & 0 & $<0.016$ & $<0.016$ & $1.00$ & \checkmark \\
GW241006\_015333 & O4b & 45 & 17.6 & 17.6 & 6.8 & 41 & $0.238$ & $0.424$ & $0.98$ & \checkmark \\
GW241102\_124058 & O4b & 19 & 21.7 & 21.7 & 4.4 & 3 & $0.021$ & $0.021$ & $1.00$ & \checkmark \\
GW241102\_144729 & O4b & 75 & 12.7 & 12.7 & 5.7 & 5 & $0.035$ & $0.035$ & $1.00$ & \checkmark \\
GW241127\_061008 & O4b & 85 & 31.3 & 31.3 & 8.8 & 0 & $<0.016$ & $<0.016$ & $1.00$ & \checkmark \\
GW241129\_021832 & O4b & 53 & 16.7 & 16.7 & 6.4 & 3 & $0.017$ & $0.017$ & $1.00$ & \checkmark \\
GW241130\_034908 & O4b & 54 & 13.8 & 13.8 & 4.5 & 31 & $0.216$ & $0.313$ & $0.98$ & \checkmark \\
GW241225\_082815 & O4b & 98 & 19.5 & 19.5 & 8.8 & 3 & $0.017$ & $0.017$ & $1.00$ & \checkmark \\
GW250108\_152221 & O4b & 90 & 11.8 & 11.8 & 5.9 & 61 & $0.354$ & $0.697$ & $0.98$ & \checkmark \\
GW250114\_082203 & O4b & 66 & 78.6 & 78.6 & 8.8 & 0 & $<0.013$ & $<0.013$ & $1.00$ & \checkmark \\
GW250118\_170523 & O4b & 68 & 13.9 & 13.9 & 6.2 & 5 & $0.035$ & $0.035$ & $1.00$ & \checkmark \\
GW250119\_025138 & O4b & 60 & 14.5 & 14.5 & 6.9 & 5 & $0.029$ & $0.041$ & $1.00$ & \checkmark \\ \hline
\end{tabular}
\end{table*}

That published value is a network number, taken over whichever detectors were observing at the time of detection. Since this search uses Hanford and Livingston only, an HLV event network SNR overestimates the Hanford--Livingston network SNR. We therefore consider the two-detector value
available to this search, written $\rho_{\rm H1+L1}$ throughout and referred to as the H1+L1 network SNR
\begin{equation}
\rho_{\rm H1+L1} \;=\; \rho_{\rm net}\,
\frac{\sqrt{\rho_{\rm H1}^{2}+\rho_{\rm L1}^{2}}}{\sqrt{\sum_{d}\rho_{d}^{2}}},
\label{eq:snrhl}
\end{equation}
in which the published network value $\rho_{\rm net}$ sets the scale and the per-detector ratio removes the Virgo quadrature contribution. Writing the correction as
a ratio allows a search pipeline measurement to correct a parameter-estimation number, since the overall normalization, and with it the difference between the two conventions, cancels out. For an event observed
only by H1 and L1 the factor is exactly unity.

The per-detector SNRs $\rho_{d}$ are not published in the event API and were recovered from the search data products. For O4 they are tabulated directly in the GWTC-4.0/4.1/5.0 search summary files. For O3 no such table is released,
and they were parsed from the GWTC-2.1 and GWTC-3 trigger files; the reconstruction reproduces the coincidence SNR recorded in the same file exactly, and the published
network value. Where several pipelines report the same
event we take the coincidence that used the most detectors.

Source-frame total masses of the detected population span $\MTOTMIN$--$\MTOTMAX\,\Msun$ with median $\MEDMTOTMAD\,\Msun$. The faintest detection has an H1+L1 SNR of $\SNRMIN$. The fraction of events recovered with respect to the official catalogs rises with mass: $5.6\%$ below $30\,\Msun$, $20.3\%$ for $30$--$60\,\Msun$, $24.2\%$ for $60$--$100\,\Msun$, and $33.3\%$ above
$100\,\Msun$. Denominators here are the \NPOPHL{} confident events observed by H1 and L1 simultaneously, not the full \NPOP{}, as 37 events were recorded by a single LIGO detector, or by one LIGO detector and Virgo, and
no Hanford--Livingston coincident search can be performed. None of these, however, is a cWB detection either, so the same restriction applies to both columns. Restricting further to the events this search is most sensitive to, specifically events characterised by $\rho_{\rm H1+L1}>10$, raises the recovered fractions to $8.1\%$ below $30\,\Msun$, $39.8\%$ between $30$ and $100\,\Msun$ and $53.3\%$ above $100\,\Msun$ (3 of 37, 35 of 88 and 8 of 15
events), corresponding to $33.3\%$, $45.5\%$ and $53.3\%$ of absolute cWB detections in the same three bins. O3 observing run detections have higher FARs than O4 detections of comparable $\sigma_{\rm net}$ (median $0.12\,{\rm yr}^{-1}$ against $0.02\,{\rm yr}^{-1}$ in O4), as it can be seen in Fig.~\ref{fig:far}. The O3 background exhibits more loud coherent-glitch families relative to signal, so more background families outrank each O3 candidate. Remarkably, GW190521~\cite{LIGOScientific:2020iuh} is not recovered although it triggers the anomaly detection: it ranks at $\FAR=2.42\,{\rm yr}^{-1}$ under the calibrated statistic (175 louder background families in 405\,yr); this miss indicates a current limitation of the pipeline due to the ranking statistic and to glitch suppression (more is discussed in Sec.~\ref{sec:discussion}).

\subsection{Sensitive volume--time}
\label{sec:vt}

\begin{figure*}[!t]
\centering
\includegraphics[width=\textwidth]{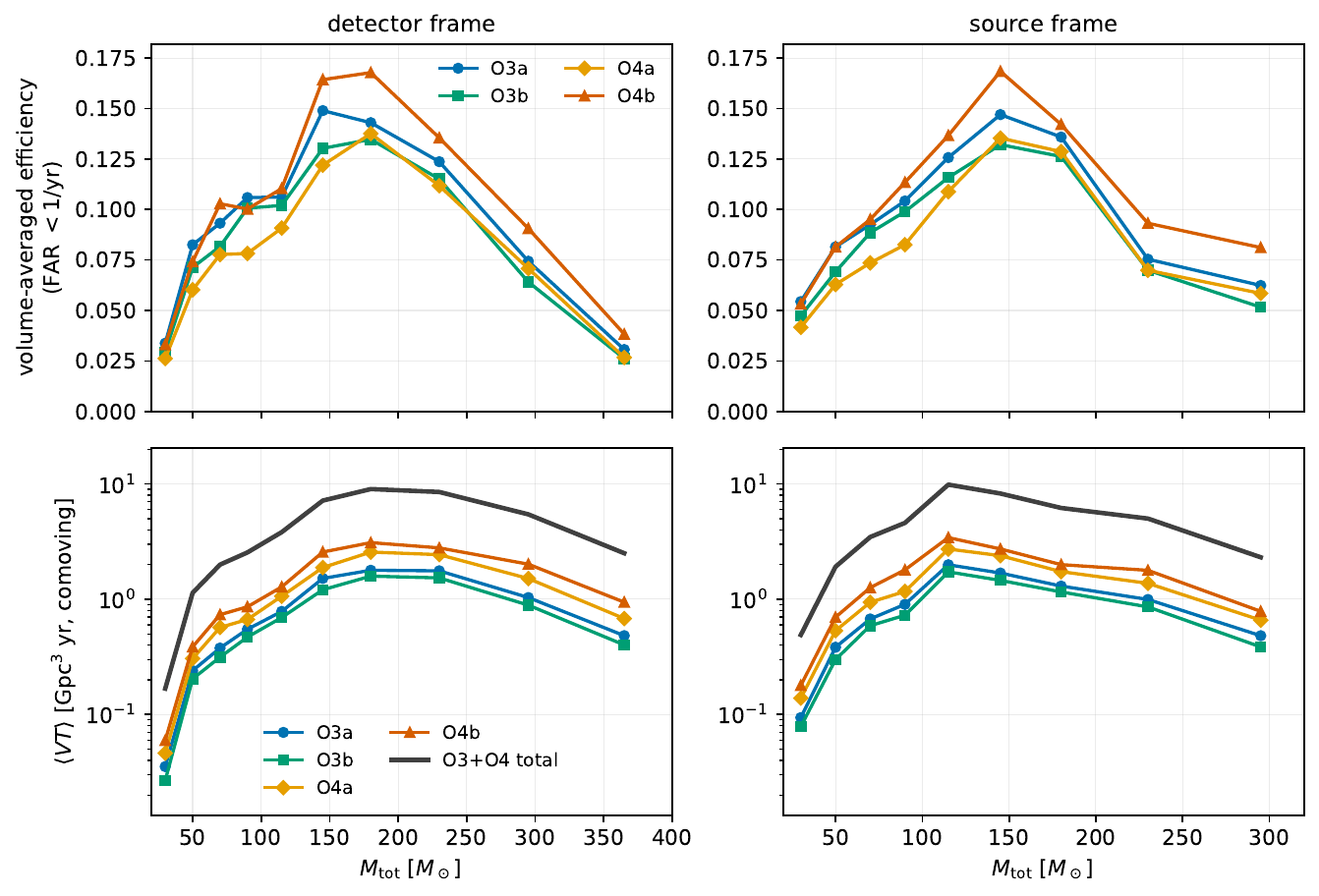}
\caption{\label{fig:vt}Left: volume-averaged efficiency of the search at
the detection criterion (calibrated $\FAR<1\,{\rm yr}^{-1}$) versus
detector-frame total mass, per observing run. Right: the corresponding
comoving sensitive volume--time $\langle VT\rangle$ versus source-frame total
mass (see text): per-injection horizons at physical network SNR $\RHOMIN$
under the run-median release spectra, comoving
$V_{\max}=\int{\rm d}V_c/(1+z)$ (flat $\Lambda$CDM, $H_0=67.9$,
$\Omega_m=0.3065$), and the source-frame rebin implied by each injection redshift. Both panels average over a population uniform in comoving volume with $(1+z)^{-1}$ rate dilation. Points whose injection support falls below $N^{\rm eff}_{\rm inj}=300$~\cite{2019RNAAS...3...66F} after the source-frame rebin are
not shown (330--400\,$\Msun$ in every run: the bank detector-frame ceiling of $400\,\Msun$ starves that source-frame bin at the horizons, $z\sim0.3$--0.5); the 260--330\,$\Msun$ points are lower bounds for the same reason.}
\end{figure*}

To convert the detection efficiency into an astrophysical sensitivity we form, per run, a sensitive volume--time (Fig.~\ref{fig:vt}). The efficiency is measured by recovering the injection campaign through the \emph{identical} adopted counting that defines the quoted FARs, applying the identical detection criterion. Injections are importance-weighted to a population uniform in
comoving volume
\begin{equation}
w_j=\rho_j^{-4}\,\Delta\rho_j\;J(z_j),\qquad
J(z)=\frac{{\rm d}V_{\max}/{\rm d}D_L}{4\pi D_L^{2}},
\label{eq:w}
\end{equation}
where $\rho_j^{-4}\Delta\rho_j$ is the Euclidean uniform volume weight of the network SNR grid level $\rho_j\in[\RHOMIN,80]$ (13 levels, each carrying
its Voronoi width $\Delta\rho_j$), following the weighted Monte Carlo construction reported in~\cite{Tiwari:2017ndi}, and the Jacobian $J$, evaluated at the injection implied luminosity distance, converts that weight to the
comoving, $(1+z)$-dilated measure of $V_{\max}$ below ($J\to1$ as $z\to0$;
median $0.5\%$--$0.6\%$ over the campaign). In a mass bin $b$,
\begin{equation}
\varepsilon_b=\frac{\sum_{j\in b}w_j\,d_j}{\sum_{j\in b}w_j},
\label{eq:eff}
\end{equation}
where $d_j\in\{0,1\}$ records whether injection $j$ satisfies the detection
criterion; injections whose network SNR falls below $\rho_{\rm min}=\RHOMIN$
are counted as undetected. Each injection is paired with the horizon of its
\emph{own} bank template (the stellar and ultramassive banks differ in
approximant and starting frequency): $D_j$ is the luminosity distance at
which that template reaches network SNR $\rho_{\rm min}$ under the run
reference spectra (leakage-free, rebuilt from the full-run strain;
Appendix~\ref{app:robust}), the same spectra that set the injections'
physical amplitudes, converted to a comoving volume
$V_{\max}(z)=\int{\rm d}V_c/(1+z)$ (flat $\Lambda$CDM,
$H_0=67.9\,{\rm km}\,s^{-1}\,Mpc^{-1}$, $\Omega_m=0.3065$, the GWTC-3
release convention). Every injection has a definite implied redshift $z_j$,
and its contribution is assigned to the source-frame mass bin containing
$\Mtot^{\rm det}/(1+z_j)$ while its normalization is taken over its
detector-frame cohort $B(j)$, so that volume migrates strictly down-mass and
the total is conserved. Collecting these, the sensitive volume--time of
source-frame mass bin $b$ is
\begin{equation}
\langle VT\rangle_b=T_{\rm obs}
\sum_{j\,:\,\Mtot^{\rm det}/(1+z_j)\in b}
\frac{w_j\,d_j\,V_{\max}(D_j)}{\sum_{i\in B(j)}w_i}.
\label{eq:vt}
\end{equation}
$T_{\rm obs}$ is the wall-clock analyzed coincident time (0.26--0.34\,yr per
run). The Monte Carlo support of a bin is quantified by the effective sample
size $N^{\rm eff}_{\rm inj}=(\sum_{j\in b}w_j)^{2}/\sum_{j\in b}w_j^{2}$, not
to be confused with the trials factor $N^{\rm eff}$ of Eq.~(\ref{eq:neff});
bins falling below $N^{\rm eff}_{\rm inj}=300$ are masked, following the
accuracy requirement for empirically measured selection functions of
Ref.~\cite{2019RNAAS...3...66F}. One convention makes these curves conservative
relative to published pipeline sensitivities: the quoted FARs inherit the
halved analyzed-time denominator of the FAR convention.

The observing runs reported in Fig.~\ref{fig:vt} differ for two reasons. First, the volume-averaged efficiency at the detection threshold peaks at $0.15$, $0.14$, $0.14$ and $0.17$ in O3a--O4b, so no run approaches unity. The curves peak at $\Mtot\approx100$--$130\,\Msun$ (O3a+O3b $\approx\VTPEAKOTHREE$, O3+O4 $\approx\VTPEAK\,{\rm Gpc}^3\,{\rm yr}$) and
turn over towards $300\,\Msun$ as in the dedicated O3 IMBH
searches~\cite{LIGOScientific:2021tfm}. Weighting $\langle VT\rangle(m)$ by the catalog
source-frame mass distribution predicts an O4 share of $63\%$ against the
observed $34/47=72\%$ ($P=0.42$), and an O3a/O3b ratio of 1.06 against
$9/4$ observed ($P=0.42$): O3a returns 1.9 its predicted share, an insignificant excess whose sign matches the residual discussed below.

\section{Comparison with existing searches}
\label{sec:comparison}

As previously mentioned, during O3a-to-O4b observing runs the published catalogs contain \NPOP{} confident detections (44, 35, 86, and 104 in O3a--O4b;
$p_{\rm astro}\geq0.5$ for GWTC-2.1/3, $\FAR<1\,{\rm yr}^{-1}$ for GWTC-4.0/5.0)~\cite{LIGOScientific:2021usb,KAGRA:2021vkt,LIGOScientific:2025slb,LIGOScientific:2026wfs}. Of these, \NCWB{} (45\%)
were identified by cWB at $\FAR<1\,{\rm yr}^{-1}$~\cite{Mishra:2022ott} (15, 9, 44,
and 52 per run; for O3a we use the cWB results of GWTC-2~\cite{LIGOScientific:2020ibl}, since
the GWTC-2.1 reanalysis did not rerun cWB). \textbf{MADGRAV} \NDET{} detections correspond to 17\% of the catalog population and 39\% of the cWB count. \NMADCWB{} of the \NDET{} are shared with cWB: the \textbf{MADGRAV} detection set is a subset of cWB, with the exclusion of 3 events. GW240406\_062847, the
only detection without a published total mass; the other two, GW190513\_205428 and GW200129\_065458, are moderate-mass binaries recovered here that cWB did not report below $1\,{\rm yr}^{-1}$. These fractions use the full published
catalogs as denominator, and are conservative lower bounds on the per-segment efficiency for two independent reasons: the catalogs include events outside the scanned Hanford--Livingston coincident segments (230 of the \NPOP{}
confident events fall within them), and they include the 37 events that have no H1+L1 coincidence at all and that no coincident search could reach (Sec.~\ref{sec:results}). The mass-binned fractions of Sec.~\ref{sec:results} remove the second of these; the comparison here is left on the full catalogs so that it can be read against the published
per-pipeline counts without restriction.

\begin{figure*}[!t]
\centering
\includegraphics[width=\textwidth]{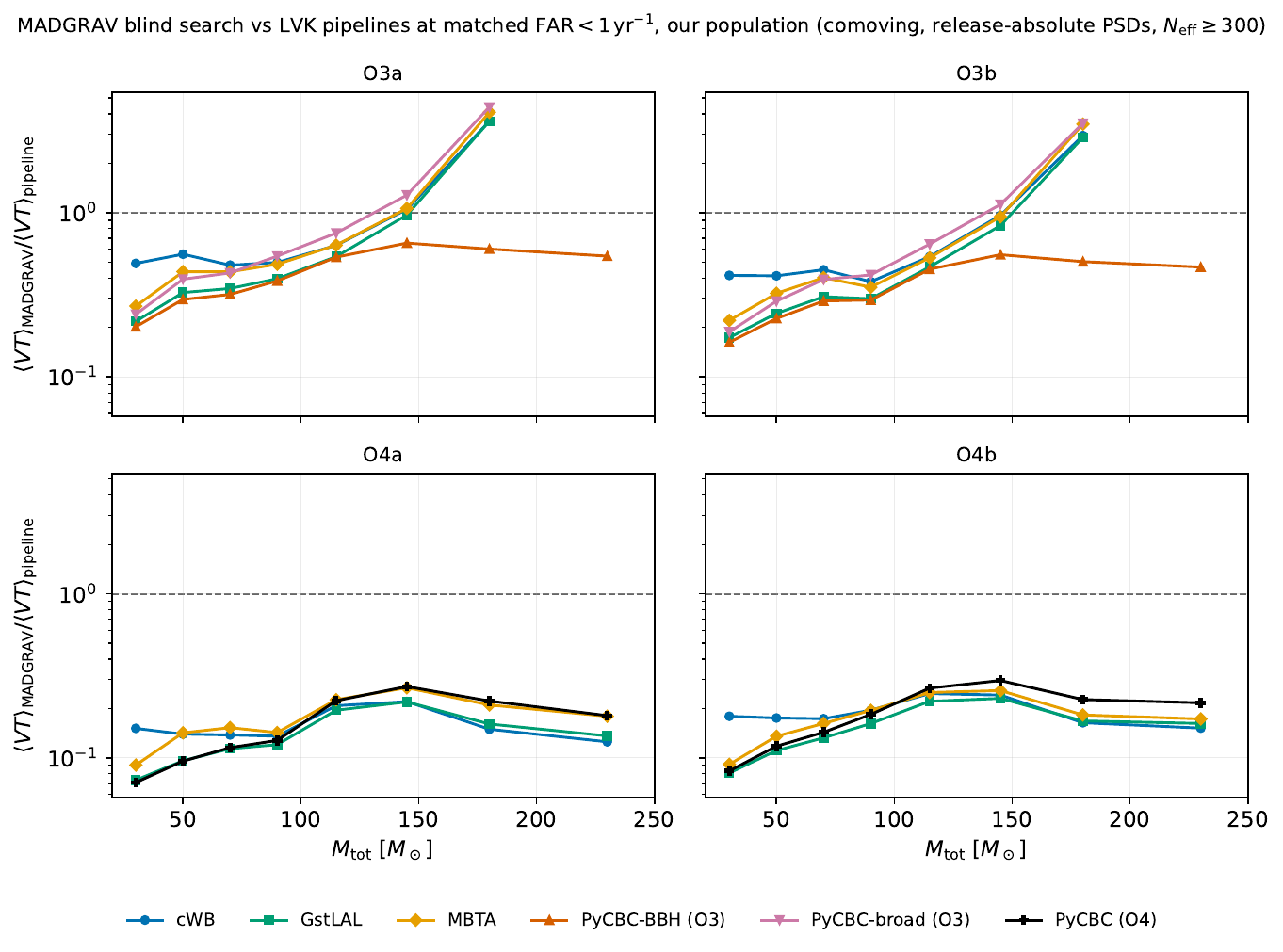}
\caption{\label{fig:fourepoch}Ratio of \textbf{MADGRAV} comoving $\langle VT\rangle$
to the LVK pipelines' at matched $\FAR<1\,{\rm yr}^{-1}$, per run, versus
source-frame total mass. Both sides are injection-set integrals on the same
population and exposure: the numerator from the campaign of
Sec.~\ref{sec:method}, the denominator from the GWTC-3 and GWTC-5.0
sensitivity-injection releases~\cite{gwtc3inj,gwtc5inj} reweighted to
it~\cite{Essick:2025zed}. Points with fewer than $N^{\rm eff}_{\rm inj}=300$
reweighted injections are dropped, which sets both limits: the releases lose
support above $230\,\Msun$, and below $20\,\Msun$ the source-frame rebin
migrates volume down-mass on the \textbf{MADGRAV} side alone.}
\end{figure*}
A sensitivity comparison at matched threshold is shown in Fig.~\ref{fig:fourepoch}. The pipeline sensitivities are
not taken from published tables: they are derived from the LVK sensitivity-injection releases for O3~\cite{gwtc3inj} and
O4~\cite{gwtc5inj} at exactly our conventions (population, mass frame, cosmology, volume law, FAR) using the
importance-reweighting estimator described with the
releases~\cite{Essick:2025zed}, and the estimator is closure-validated by reproducing each release own surveyed $\langle VT\rangle$ under the release conventions to $2\times10^{-5}$. Both sides weight the population uniformly
in comoving volume with $(1+z)^{-1}$ rate dilation (Sec.~\ref{sec:vt}). The published external benchmarks, the O3 IMBH search~\cite{LIGOScientific:2021tfm}, are read from those works at their own conventions and not recomputed here. In O3a the \textbf{MADGRAV} reaches 0.20--0.54 of the matched filtering pipelines and 0.48--0.56 of cWB sensitivity in 20--100\,$\Msun$, rising to 0.54--1.28 and 0.63--1.05 at 100--160\,$\Msun$;
in O3b the ratios are 0.16--0.45 below $100\,\Msun$ and 0.45--1.12 at 100--160\,$\Msun$, closing most of the gap to matched filtering pipelines and cWB at the top of the mass range where the search is designed to operate. The 160--200\,$\Msun$ bin shows ratios of 2.9--4.4 against every pipeline except PyCBC-BBH (0.50--0.60. The count ratio against cWB exceeds the corresponding $\langle VT\rangle$ ratio below $100\,\Msun$, where most catalog events lie,
in O4 and marginally in O3: O3a $9/15=0.60$ against 0.48--0.56, O3b $0.44$ against
0.38--0.45, O4a $0.33$ against 0.14--0.15, and O4b $0.35$ against
0.17--0.20. This ordering is expected rather than anomalous: the two
quantities are conditioned differently. The count ratio is evaluated on
catalog events---that is, conditioned on the event having been detected by
some pipeline---a population with median network SNR $9.9$--$10.6$ in O4,
whereas $\langle VT\rangle$ integrates an unconditioned astrophysical
population whose comoving volume weighting places its median at SNR $7.0$.
The search efficiency differs sharply between those regimes: for
$\Mtot\geq20\,\Msun$ it is below $0.01$ at $\rho_{\rm net}=8$,
$0.05$--$0.09$ at $10$, $0.17$--$0.26$ at $12$, $0.42$--$0.54$ at $15$ and
$0.68$--$0.76$ at $20$. A search whose SNR threshold is higher than its
comparators must therefore rank better on catalog counts than on
volume-averaged sensitivity, and the gap is wider in O4 because the template
pipelines reach at SNR $\sim7$ is greater there.

Applied directly to the catalog population the two measures are in agreemnt: summing
this search per-cell recovery probability over the comparable catalog
events of each run (Table~\ref{tab:cmp}) predicts 7.8, $4.0$, $11.6$ and $23.2$
detections in O3a O3b, O4a and O4b against $9$, $4$, $15$ and $18$ found, the ratios are
$0.86$, $1.00$, $0.77$ and $1.29$. In O4 the search sits at
0.07--0.20 of the pipelines below $100\,\Msun$, peaks at 0.20--0.30 at
100--160\,$\Msun$, and declines to 0.13--0.22 by 200--260\,$\Msun$. With the
reference spectra corrected (Sec.~\ref{sec:vt}) the horizon is restored while
the efficiency term is renormalized in the opposite direction, so this
deficit is a property of the search itself and not solely of its spectra.
The GWTC-5.0 release provides a single PyCBC search, plotted as such in the
O4 panels of Fig.~\ref{fig:fourepoch}. Fig.~\ref{fig:fourepoch} uses the
comoving, source-frame numerator of Fig.~\ref{fig:vt}, so the two are
mutually consistent; the mass-ceiling lower-bound caveat on the
260--330\,$\Msun$ bin applies to both.

\section{Discussion and outlook}
\label{sec:discussion}

We have demonstrated that an anomaly-detection-based deep learning pipeline, trained on O3a data and frozen before testing it on O3a-to-O4b observing runs, detects \NDET{} binary black hole mergers at a calibrated $\FAR<1\,{\rm yr}^{-1}$. The search returned zero detections without a catalog counterpart, indicating not only the capability to detect events, but also its ability to suppress environmental or instrumental transient noise that might mimic real events. 

Most of the events that \textbf{MADGRAV} does not recover are missed due to different aspects. Of the $139$ confident catalog events with $\Mtot>10\,\Msun$ observed in Hanford--Livingston coincidence at $\rho_{\rm H1+L1}>10$, this search recovers \NDET{}; we have investigated each of the 92 missed detections. None is explained by data coverage, in fact all segments containing events have been scanned by the pipeline. Of the 92 events missed, 62 (corresponding to $67\%$) never produce a $\sigma_{\rm net}>4$ trigger. The trigger statistic sums two per-detector anomaly scores, each of which floors near zero below single-detector SNR $\sim7$--$8$, and these events are typically asymmetric, with one detector well above threshold and the other near zero; 24 events (corresponding to $26\%$) produce a trigger but are then lost in the ranking, falling below the $\ln\Lambda=4.0$ candidate floor. Of the remaining 6, 2 are either rejected as glitches or scored with FAR above threshold (such as GW190521).

We can also identify classes of missed events based on their total source-frame mass. Below $30\,\Msun$ the loss is almost entirely at the trigger level, in fact, 29/32 never fire the trigger. Between $30$ and $100\,\Msun$ the two causes are comparable, $31$ don't fire the trigger and $20$ suffer from the ranking statistics. Above $100\,\Msun$ the trigger is the minority cause, $2$ of $7$, and the ranking statistic is the main cause for outranking signal candidates. This suggests that next version of \textbf{MADGRAV} will have to include multiple modifications to improve its efficiency over a broader region of total source-frame masses and SNR.

The ranking-limited class is not a glitch-gate problem, and lowering the candidate floor would not necessarily recover its events. Of the 77 events that fire the trigger, 53 are above $\ln\Lambda=4.0$ threshold and 48 of those pass the CNN gate: the
gate rejects most of the background triggers it sees, but only two real catalog events. Lowering the threshold to admit the 23 ranking-limited events above $30\,\Msun$ would also require regenerating the time-slide background with the same lowered threshold, which in turn would increase the FAR, as a factor $2$--$3$ more background triggers at $\ln\Lambda\geq3$, for example. 

We conclude that for an updated version of \textbf{MADGRAV} several components could be improved. First, the trigger sums two independently floored per-detector scores; combining the detectors before thresholding is the change that would act on the two thirds of the misses that never trigger. Second an improved CNN stage using not only real data, but also a set of simulated glitches can help suppress the glitch contamination and maintaining a lower FAR. Various latent space compressions have been explored, and although most levels of compression, including the overcomplete case, reach very similar sensitivities, as shown in Appendix~\ref{app:bottleneck} some optimisation per mass region might help recovering more events, particularly in the lower mass region. 

A possible further enhancement of the pipeline sensitivity can come from the features extracted during the pipeline workflow that contribute to $\ln\Lambda$: these could be combined in a gradient algorithm to find the ideal hyperplane that allows optimal separation between real signal and glitches or noise, as done for example in~\cite{Mishra:2022ott}.

Based on the evidence presented here, we think that anomaly-detection searches for gravitational waves are not anymore just a proof of concept but they are an independent,
calibrated detection channel for the massive BH binary mergers.

\section*{Declaration of competing interest}
The authors declare that they have no known competing financial interests or personal relationships that could have appeared to influence the work reported in this paper.

\section*{Data availability}
The strain data analysed in this work are public and available from the Gravitational Wave Open Science Center (gwosc.org)~\cite{KAGRA:2023pio,LIGOScientific:2025snk,LIGOScientific:2026jgl}. The code needed to reproduce the results is available in the public \textbf{MADGRAV} repository~\cite{madgrav_repo}.

\section*{Acknowledgements}
We thank the LVK bursts group and all-sky-short subgroup for useful feedback during the preparation of this work. The computational results have been achieved using the Austrian Scientific Computing (ASC) infrastructure, including the Vienna Scientific Cluster (VSC-5), the MUSICA (Multi-Site Computer Austria) system, and the project \"OAW-MUSICA-2026. This research has made use of data or software obtained from the Gravitational Wave Open Science Center (gwosc.org)~\cite{KAGRA:2023pio,LIGOScientific:2025snk,LIGOScientific:2026jgl}, a service of the LIGO Scientific Collaboration, the Virgo Collaboration, and KAGRA. This material is based upon work supported by NSF's LIGO Laboratory which is a major facility fully funded by the National Science Foundation, as well as the Science and Technology Facilities Council (STFC) of the United Kingdom, the Max-Planck-Society (MPS), and the State of Niedersachsen/Germany for support of the construction of Advanced LIGO and construction and operation of the GEO600 detector. Additional support for Advanced LIGO was provided by the Australian Research Council. 
Virgo is funded, through the European Gravitational Observatory (EGO), by the French Centre National de Recherche Scientifique (CNRS), the Italian Istituto Nazionale di Fisica Nucleare (INFN) and the Dutch Nikhef, with contributions by institutions from Belgium, Germany, Greece, Hungary, Ireland, Japan, Monaco, Poland, Portugal, Spain. KAGRA is supported by Ministry of Education, Culture, Sports, Science and Technology (MEXT), Japan Society for the Promotion of Science (JSPS) in Japan; National Research Foundation (NRF) and Ministry of Science and ICT (MSIT) in Korea; Academia Sinica (AS) and National Science and Technology Council (NSTC) in Taiwan.

\section*{Declaration of generative AI and AI-assisted technologies in the manuscript preparation process}
We acknowledge the following uses of generative AI tools in preparing this manuscript. Claude Code was used to support the organisation of the structure and documentation of the public code repository, to produce the pipeline flowchart in Fig.~\ref{fig:madgrav-flow} from an author written description, and to diagnose formatting problems in the bibliography style file. Grammarly was used for spelling and grammar checks of author written text. No AI tool was used to generate scientific content, data, results, or references, and the authors take full responsibility for the content of the manuscript.

\appendix

\section{Different latent space compressions}
\label{app:bottleneck}

To test whether the detections depend on the overcomplete latent
(Sec.~\ref{sec:architecture}) or on the full training procedure, we retrained the CAE on an independent draw of 32 of the 143 O3a blocks ($36.4$\,h per detector) with the same signal bank, recipe and hyperparameters, with the addition of a bottleneck layer of $k$ channels at the narrowest layer (two $1\times1$
convolutions, $128\to k\to128$; latent $k\times512$ dimensions) for
$k=32$, 20, 13 and 1, i.e.\ compression factors 4, 6.4, 9.8 and 128 of the model presented in the paper. All models, the frozen one included, were re-standardised on a common sample of 3000 O3a noise windows per detector and compared at matched FAR. Each model threshold reproduces the frozen model fraction of $9\times10^{6}$ random Hanford--Livingston noise pairs above $\sigma_{\rm net}=4$ (the model operating point). The comparison is at the trigger level.

At the matched threshold every margin-trained model down to $k=13$ triggers on all 47 detections, on the near-miss GW190521\_030229, and on the same 83 catalogue events as the frozen model (Table~\ref{tab:bottleneck_events};
$k=32$ adds one below-floor event, GW230609\_064958). The unsupervised stage alone triggers on at most 3 of the 136 catalogue events at any width.

\begin{table}[!htb]
\centering
\setlength{\tabcolsep}{4pt}
\caption{Catalog events above the matched threshold: the 47
detections, the near-miss GW190521, and all 136 scorable GWTC events ($M_{\rm tot}>10\,\Msun$, H--L coincident, $\rho_{\rm HL}>10$).}\label{tab:bottleneck_events}
\begin{tabular}{lrrccr}
\hline
Model & $k$ & latent dim. & detected & GW190521 & all \\
\hline
frozen     & 128 & 65,536 & 47/47 & yes & 83 \\
bottleneck & 32  & 16,384 & 47/47 & yes & 84 \\
bottleneck & 20  & 10,240 & 47/47 & yes & 76 \\
bottleneck & 13  & 6,656  & 47/47 & yes & 83 \\
bottleneck & 1   & 512    & 24/47 & yes & 38 \\
\hline
\end{tabular}
\end{table}

\begin{figure*}[!t]
\centering
\includegraphics[width=\textwidth]{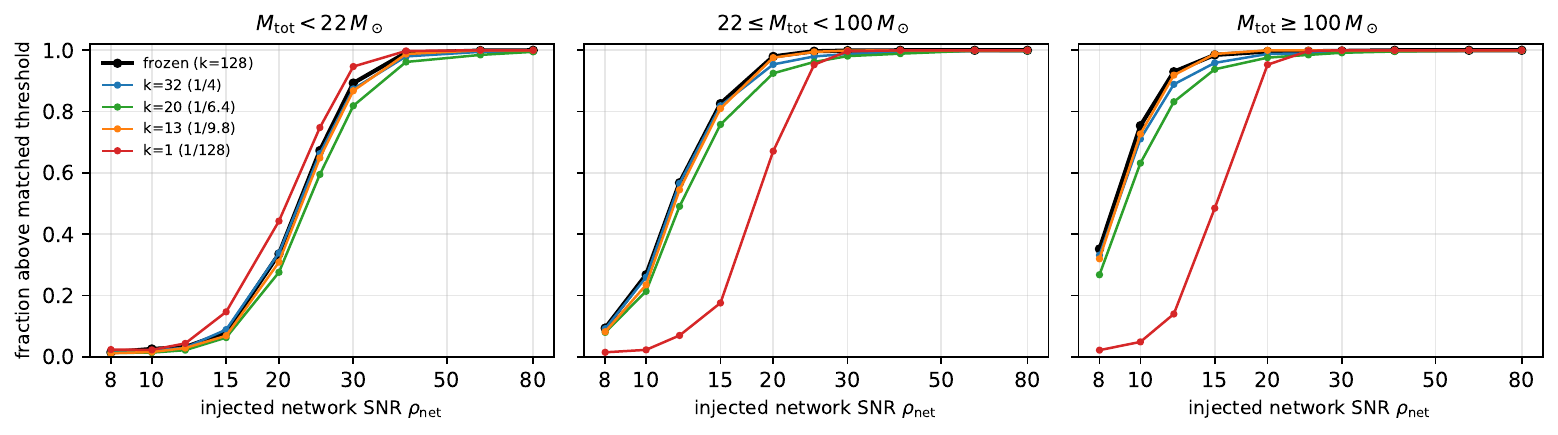}
\caption{Trigger-level injection recovery versus network SNR per
detector-frame mass stratum at the production operating point. Black:
frozen model; colours: bottleneck widths $k$.}
\label{fig:bottleneck_recovery}
\end{figure*}

We study the same $64{,}350$ seeded injections of the campaign of
Sec.~\ref{sec:method} on all compression models and show the results in Fig.~\ref{fig:bottleneck_recovery}. For $k=32$ the recovered fraction is within one percentage point of the frozen
model overall and within two in every mass range, identical from
$\rho_{\rm net}=60$, with the only deficit, 4 points, at
$\rho_{\rm net}=10$--12 for $M_{\rm tot}\geq100\,\Msun$. For $k=20$ the model is 3 percentage points less efficient overall. At $k=13$ the overall fraction still matches, but the noise tail is much heavier: a noise-pair tail of $10^{-4}$ requires $\sigma_{\rm net}=9.6$, against 6.1 for the frozen model and 5.4 for $k=32$. At $k=1$ the model recovers nothing below $\rho_{\rm net}=20$ for $M_{\rm tot}>22\,\Msun$, but it interestingly exceeds the frozen model below $22\,\Msun$, which suggests that a mass-targeted model or ensemble is a natural
extension. 

We conclude that a compression of the latent space dimensions up to a factor of 4 does not impact on the number of detections or on the overall efficiency of the model; such compression could easily be adopted. Beyond a compression factor of 6 it reduces the efficiency, and beyond a factor 10 it affects the dynamic range. However it should be mentioned that the separation between signals and noise comes from the margin objective and not from the latent space size, in fact the same architecture trained jointly from the first epoch, or with the unsupervised stage alone, does not recover the catalog events.

\section{Validation and calibration checks}
\label{app:robust}

\textbf{Reference spectra.} The reference spectra that set injection
amplitudes and horizons (Sec.~\ref{sec:vt}) are rebuilt for every run from
full-run GWOSC strain with a 4\,s transform (0.25\,Hz resolution). They
reproduce the published BNS inspiral ranges of all four runs to 0.94--1.06.

\textbf{Systematics of $\langle VT\rangle$.} Efficiencies on 25 event-free
segments per run relative to the event-hosting segments are 1.50
(20--40\,$\Msun$) falling to 1.00 (above 200\,$\Msun$) in O3a, 0.88--0.99
in O3b and 1.00--1.13 in O4, so no correction is applied. The per-segment
efficiency spread gives a $\langle VT\rangle$ systematic of
$\sigma_{\rm rel}=0.62$ (20--40\,$\Msun$) falling to 0.04 (330--400) in O3a
and 0.21 to 0.01 in O3b. 

\textbf{Empirical null calibration.} The pseudo-foreground construction of
Sec.~\ref{sec:method}, applied to all gate-passing background families at
$x=1\,{\rm yr}^{-1}$, gives $\mathcal{K}=5.60$ (O3a), $4.59$ (O3b), $1.43$
(O4a) and $3.62$ (O4b) against the foreground-excluded background, with the glitch-arm gate applied to the pseudo-foreground population and to its background as to the candidates (without the gate: $5.54$, $4.54$, $2.34$, $5.46$; the gate removes the glitch pairs that dominated the O4 pseudo-foreground, so $\mathcal{K}$ drops where the background was glitch-dominated and is unchanged in O3). We performed three checks to establish what $\mathcal{K}$ measures. A single-channel unconditioned rank returns $\mathcal{K}=1.00$ in all four runs. We also find that only $2$ of $6\,624$ passing background families lie within $\pm4$\,s of an undetected confident catalog event. The minimum over the two channels implies a factor $1.75$--$1.94$; the arm-conditioned counting supplies the remainder and varies per event ($0.7$--$265$), so $\mathcal{K}$ is quoted per run.

\textbf{Zero-lag census.} Four unmatched zero-lag triggers had
$\FAR<1\,{\rm yr}^{-1}$ before the consistency veto (three in O3b, one in
O4b); all four were removed by it. After the veto, two unmatched triggers
remain, both in O3a, at 2.4 and $4.7\,{\rm yr}^{-1}$, against $\approx3$
expected below $5\,{\rm yr}^{-1}$ in 0.61\,yr of analyzed time, and none
below $1\,{\rm yr}^{-1}$ against $\approx0.6$ expected---consistent with a
calibrated statistic.

\textbf{Foreground excision.} The time-slide background of a detected event can be constructed either containing its own windows paired with other slides or excluding every background pair within $\pm4$\,s of a detection in its segment. Table~\ref{tab:detections} reports both FARs: the two agree to
$5\times10^{-5}\,{\rm yr}^{-1}$ for all but eight events, $\mathcal{K}$
moves by less than$0.61$ (inclusive-background values under the gate: $5.58$, $4.61$, $2.03$, $4.04$ against exclusive $5.60$, $4.59$, $1.43$, $3.62$; the largest move is O4a, where the gate leaves few glitch families in the pseudo-foreground) in every run, and one O4b detection (GW240406\_062847) rises above $1\,{\rm yr}^{-1}$ under the inclusive background.

\section{Construction of $p_{\rm astro}$}
\label{app:pastro}

The per-event probability of astrophysical origin quoted alongside the
primary FAR is a two-component Poisson mixture (FGMC), fitted per run in
analogy to the cWB method of GWTC-4.0~\cite{2015PhRvD..91b3005F,LIGOScientific:2025zdk}. The fit
variable is the per-arm foreground-excluded FAR
$x=\FAR_{\rm excl}$ before the calibration factor $\mathcal{K}$, on
$x\in[0,x_{\rm max}]$ with $x_{\rm max}=1\,{\rm yr}^{-1}$; $\mathcal{K}$ enters only through the admission rule, applied identically
to candidates and injections. With $\Lambda_s$, $\Lambda_n$
the expected signal and noise counts and $p_s$, $p_n$ their densities on
$[0,x_{\rm max}]$, the extended likelihood of the observed $\{x_i\}_{i=1}^{N}$ and the
per-event probability are
\begin{equation}
\begin{aligned}
  \mathcal{L}(\Lambda_s,\Lambda_n)\;&=\;
  e^{-(\Lambda_s+\Lambda_n)}\prod_{i=1}^{N}
  \bigl[\Lambda_s\,p_s(x_i)+\Lambda_n\,p_n(x_i)\bigr],
  \\
  p_{\rm astro}(x_i)\;&=\;
  \frac{\Lambda_s\,p_s(x_i)}{\Lambda_s\,p_s(x_i)+\Lambda_n\,p_n(x_i)}\,.
\end{aligned}
  \label{eq:fgmc-like}
\end{equation}

The FAR of a noise candidate is uniform, $p_n(x)=x_{\rm max}^{-1}$, and $\Lambda_n=x_{\rm max}\,T_{\rm an}$, with $T_{\rm an}$ the analyzed coincident exposure (half the wall-clock livetime of Sec.~\ref{sec:vt}, since each 4\,s window contributes a 2\,s analyzed tile).
Setting $\Lambda_n$ to its null expectation makes the estimate
self-calibrating (a free $\Lambda_n$ would be unconstrained by $4$--$19$
candidates per run); only $\Lambda_s$ is fitted.

We measure the signal density $p_s$ using the per-run injections of
Sec.~\ref{sec:method} that pass through the identical chain, gate, veto and per-arm counting and that produce the candidate FARs. $p_s$ is the histogram of $x$ over injections and following the full detection criterion, weighted uniform-in-comoving-volume (Sec.~\ref{sec:vt}) and reweighted in total mass to the published catalog distribution; intervals are 5--95\% ranges over 200 bootstrap replicates.
The pinned noise counts are $\Lambda_n=0.146$, $0.132$, $0.172$ and $0.156$
in O3a--O4b, versus the fitted $\Lambda_s=8.98$, $3.99$, $14.91$ and $18.89$:
the mixture is signal-dominated by two orders of magnitude.

We test the construction in two ways. First, we let the noise count $\Lambda_n$ float instead of pinning it to the time-slide null expectation. The fit returns $1.16$ in O3a and drives $\Lambda_n$ to zero in the other three runs, where a few candidates cannot constrain it; the lowest $p_{\rm astro}$ of the \NDET{} detections changes from 0.913
to 0.970. Second, we multiplied the noise count by the calibration factor,
$\Lambda_n\to\mathcal{K}\Lambda_n$, as a conservative choice.
42 detections stay above $0.95$; the two lowest, GW240406\_062847 (O4b),
and GW230707\_124047 (O4a), drop to $0.83$ and $0.88$. No quoted
value uses this limit. Two properties of the values follow from the
construction. At fixed fit, $p_{\rm astro}$ decreases monotonically with
$\FAR$, so it never reorders candidates and adds no detection beyond the
FAR threshold. Where the signal density reaches the floor of the injection
campaign, the value is an upper bound; this is the case for
GW190521\_030229 ($p_{\rm astro}<10^{-3}$).

\bibliographystyle{elsarticle-num}
\bibliography{madgrav}

\end{document}